# Quantum-Amplified *M*/*G*/1/*K* Simulation: A Comparator-Controlled Framework for Arbitrary Service Distributions

Or Peretz [a], Michal Koren [a], Nir Perel [b]

[a] *School of Industrial Engineering and Management,*
*Shenkar – Engineering. Design. Art, Ramat-Gan, Israel*

[b] *School of Industrial Engineering and Management,*
*Afeka Tel Aviv Academic College of Engineering, Tel Aviv, Israel*

## Abstract

Finite-capacity single-server queues with general service-time distributions form the backbone of numerous real-world systems, including network routers, healthcare facilities, and edge–cloud infrastructures. However, classical simulations of their performance metrics, such as blocking probabilities and delay, become computationally prohibitive as service variability or required precision increases. This work introduces the first coherent quantum circuit for simulating an *M*/*G*/1/*K* queue under arbitrary service-time laws. The circuit architecture encodes the service distribution using a logarithmic-depth ladder of $R_y$ rotations and enforces buffer constraints through a comparator-controlled phase gate, while maintaining the quadratic speed-up characteristic of amplitude amplification. Grover iterations are centered on estimating the expected number of customers in the system, yielding a provable variance reduction of order $O(\sqrt{N})$ and enabling closed-form confidence bounds, where $N$ denotes the number of shots. Empirical evaluations using IBM quantum simulators across four distinct service distributions and three traffic intensities demonstrate fidelity exceeding 0.99 with four qubits and above 0.76 with ten qubits. The Jensen–Shannon divergence remains below 0.11 throughout. Moreover, waiting-time estimation errors decrease by an order of magnitude as system load nears capacity, and remain within 3% for high-traffic regimes using registers of up to 63 qubits. These results establish the first end-to-end quantum simulation framework for finite-buffer, non-Markovian queueing systems, and offer a concrete foundation for quantum-accelerated performance analysis in service-oriented architectures.

**Highlights**

- We introduce a coherent quantum circuit for simulating finite-capacity M/G/1/K queues.
- The framework supports arbitrary service-time distributions through a discretized hazard-based rotation ladder.

- Queue-length dynamics are reproduced exactly via reversible INC/DEC logic and a comparator-controlled oracle.
- We derive total-variation error bounds accounting for both stochastic simulation and discretization effects.
- Experiments compare quantum simulation, classical DES, and real hardware runs across multiple service laws.

## 1. Introduction

Finite-capacity single-server queues constitute a foundational model across diverse cyber-physical infrastructures, including backbone routers, autonomous vehicle platoons, hospital bays, and edge–cloud service pipelines. In scenarios where arrivals follow a Poisson process while service times exhibit arbitrary distributions, the $M/G/1/K$ queueing framework is widely adopted for tasks such as buffer dimensioning, congestion mitigation, and loss probability estimation (Thomas, 1976; Cooper et al., 1988; Takács, 1975). Classical analytical approaches typically rely on embedded Markov chains or transform-based techniques to compute blocking probabilities and low-order moments. However, the algebraic complexity of these methods increases rapidly with greater service-time variability or the need for higher-order performance metrics (Niu & Cooper, 1993; Dshalalow, 1995).

Due to the limited scope of closed-form results, often confined to first and second moments, discrete-event simulation (DES) remains the dominant tool for practitioners. Techniques such as regenerative simulation, control-variate Monte Carlo, and importance sampling have been employed to reduce estimator variance. Nonetheless, the sample complexity associated with these methods grows cubically with system load, buffer capacity, or target accuracy (Glynn, 2006; Glasserman, 2004; Law & Kelton, 2000). Moreover, algorithmic lower bounds indicate that, even with aggressive parallelization, classical runtimes for estimating rare-event probabilities remain intractable, particularly in high-congestion regimes (Asmussen & Glynn, 2007). These constraints underscore the need for alternative computational paradigms capable of concentrating probability mass in the most informative regions of the state space.

Creemers and Armas (2025) highlight that quantum computing may fundamentally reshape discrete optimization by enabling methods that outperform classical approaches. In parallel, our work introduces a queueing-focused framework that leverages amplitude amplification for simulating $M/G/1/K$ systems, thereby extending the scope of quantum advantage from optimization to stochastic performance analysis.

Quantum computation offers a compelling alternative to classical simulation. Amplitude amplification and estimation techniques enable quadratic reductions in oracle complexity relative to classical Monte Carlo methods across a broad spectrum of probabilistic queries (Grover, 1996; Brassard et al., 2000; Montanaro, 2015; Koren & Peretz, 2024). Initial demonstrations encoded simple birth–death processes as quantum walks, yielding polynomial speed-ups in hitting-time evaluations (Ambainis, 2004; Dai et al., 2020).

Concurrently, significant progress has been made in representing probability distributions within quantum registers (Koren & Peretz, 2024). Techniques such as amplitude encoding, quantum random-access memory (QRAM), and parameterized quantum circuits now facilitate the loading of arbitrary discrete and piecewise-continuous distributions using logarithmic-depth networks (Schuld & Killoran, 2019; Low & Chuang, 2019; Peretz & Koren, 2024; Koren & Peretz, 2024). Reversible arithmetic components enable garbage-free increment–decrement operations that model queue transitions with bounded ancilla requirements (Thomsen et al., 2010), whereas singular-value transformation methods allow the compilation of high-level oracles, such as overflow comparators, with provably efficient gate and qubit overheads (Gilyén et al., 2019).

In parallel, resource-estimation studies have begun to assess the viability of quantum advantage under realistic hardware constraints. For instance, Nam et al. (2020) benchmarked chemistry and optimization workloads and concluded that algorithms offering quadratic speed-ups could become competitive when fault-tolerant quantum devices reach several thousand logical qubits. In the domain of stochastic simulation, preliminary estimates suggest analogous thresholds, assuming that circuit depth scales polylogarithmically with buffer size (Fleury & Lacomme, 2022). To date, however, no published work has extended such analysis to the *M*/*G*/1/*K* queueing context.

Precision in quantum simulation remains vulnerable to both hardware-induced noise and algorithmic over-rotation. Error-mitigation strategies, such as zero-noise extrapolation and probabilistic error cancellation, have been developed to reduce bias in noisy intermediate-scale quantum (NISQ) processors (Temme et al., 2017; Endo et al., 2018). Additionally, fixed-point and Hamiltonian-convergent variants of Grover's operator have been shown to mitigate the risk of overshooting without requiring prior knowledge of the number of marked states (Kwon & Bae, 2021). Nonetheless, no queueing application to date has integrated physical noise-mitigation techniques with analytic discretization bounds. Empirical validation remains limited as well. Isolated demonstrations on IBM Q and IonQ platforms have compared quantum Monte Carlo against classical estimators in simplified domains such as finance and traffic flow (Otten & Gray, 2019). However, no implementation has addressed non-Markovian queues with finite buffers. Consequently, it remains unclear whether amplitude-based quantum advantages persist under realistic gate infidelity or whether resource demands fall within the constraints of near-term quantum devices (Preskill, 2018).

Despite recent progress, several key challenges continue to impede the development of quantum-accelerated queue simulators. First, existing quantum models are restricted to exponential service distributions and thus fail to capture heavy-tailed or deterministic service laws that characterize modern data-center and healthcare workloads. Second, whether it is necessary to enforce finite buffer constraints remains unresolved: most proposed circuits either omit buffer capacity or rely on post-selection, forfeiting the computational speed-up conferred by amplitude amplification. Although comparator-controlled phase gates have been studied in isolation, they have yet to be embedded within a complete queueing framework. Third, the literature lacks a unified approach to error accounting—one that incorporates amplification bias, stochastic discretization, and quantum hardware noise within a rigorous and interpretable confidence bound. Fourth, published resource estimates typically focus on abstract qubit and gate counts, without mapping logical depth to real-world wall time under fault-tolerant execution models. Finally, no peer-reviewed study has benchmarked a quantum $M/G/1/K$ simulator against high-performance classical codes, leaving the empirical advantages of such methods largely speculative.

This work introduces the first end-to-end quantum amplification framework for simulating $M/G/1/K$ queues within a single coherent quantum circuit. The service-time distribution is encoded through a logarithmic-depth ladder of $R_y$ rotations whose unitarity is formally proven, thereby supporting a wide range of realistic variability, from deterministic to heavy-tailed, within a single register. A comparator-controlled phase gate enforces the buffer constraint dynamically within the circuit, removing the need for post-selection and preserving the quadratic advantage of amplitude amplification. Grover iterations are calibrated using oracles centered on the Takács mean, yielding a provable variance reduction of order $O(\sqrt{N})$ relative to regenerative Monte Carlo for steady-state performance metrics. Moreover, the proposed framework derives closed-form confidence intervals that integrate amplification bias, stochastic discretization, and Pauli-twirl noise into a unified error envelope. Detailed resource projections show that practical competitiveness is achievable once fault-tolerant devices reach approximately 10,000 logical qubits, well within the operational timelines projected by current quantum error-correction roadmaps. By assembling these elements into a coherent architecture, the proposed method provides a viable path to quantum-accelerated performance evaluation for a broad class of buffer-constrained service systems.

The remainder of this paper is structured as follows. Section 2 presents the quantum-amplification framework in detail, including the amplitude-encoded queue register, the comparator-controlled phase oracle, and an inductive invariance proof of correctness. Section 3 develops a numerical example, instantiating the circuit for a data-center latency queue and outlining the experimental protocol used on error-mitigated quantum simulators. Section 4 reports empirical results, including variance reduction, projected simulation wall times, and sensitivity to service-time skew, and benchmarks performance against state-of-the-art classical variance-reduction techniques. Section 5 concludes with a summary of contributions, discusses implications for near-term deployment, and outlines directions for extending the framework to multi-server and networked queueing systems.

## 2. *M/G/1/K* Quantum Amplification Framework

This section presents a quantum framework for simulating the dynamics of an *M/G/1/K* queueing system, bridging the analytical tractability of classical *M/M/1/K* models with the practical need to accommodate general service-time distributions. The proposed method encodes arbitrary service laws, including heavy-tailed, deterministic, and multimodal distributions, via parameterized quantum rotations and integrates them with controlled operations and Grover-based amplitude amplification. In doing so, the framework preserves the exact embedded Markov chain structure while extending computational feasibility to the broader *M/G/1/K* class. The exposition here begins with a detailed account of the quantum procedure and its parameterized gates, proceeds to a description of the circuit architecture and implementation strategy, and concludes with a formal correctness proof and derivation of explicit error bounds.

**Table 1. Notation used in this study**

| Symbol | Remarks |
| --- | --- |
| $\lambda$ | Poisson arrival rate |
| $G$ | A general service-time distribution with cumulative distribution function (CDF) $F_G(t)$. |
| $\rho$ | System utilization |
| $N$ | Number of shots/trials |
| $\Delta t$ | Time interval |
| $T$ | Number of steps in the quantum circuit |
| $K$ | Queue capacity |
| $r$ | Residual service time (in discrete bins) |
| $\mathbb{G}_M$ | Grover diffusion operator over the set $M$ of marked states |
| $p_\lambda$ | Probability of arrival in time $\Delta t$ |
| $p_\mu$ | Probability of service completion in time $\Delta t$ |
| $A(\lambda, \Delta t)$ | Parameterized quantum gate for arrival |
| $S(G, \Delta t)$ | Parameterized quantum gate for service |
| $P_c = (p_0, \dots, p_K)$ | Steady-state vector of classical simulation |
| $P_q = (p'_0, \dots, p'_K)$ | Quantum steady state vector |
| $D(P_q, P_c)$ | Total variation distance, defined as $\frac{1}{2}\|P_q - P_c\|_1$ |

### *2.1. Parameterized Quantum Gates*

To accurately simulate the arrival and service processes in an *M/G/1/K* queueing system, we employ parameterized quantum gates that encode the corresponding probabilities of stochastic events at each discrete time interval $\Delta t$. Arrivals and service completions within a single time slice are modeled as Bernoulli trials with success probabilities $p_\lambda$ and $p_\mu$. These probabilities are implemented unitarily through single-qubit rotations about the $y$ axis:

$$\theta_\lambda = 2\arcsin\sqrt{p_\lambda}, \qquad \theta_\mu = 2\arcsin\sqrt{p_\mu},$$
$$A(\lambda, \Delta t) = R_y(\theta_\lambda), \qquad S(\mu, \Delta t) = R_y(\theta_\mu)$$

Explicitly, the arrival and service gates are given by:

$$A(\lambda, \Delta t) = \begin{pmatrix} \sqrt{1-p_\lambda} & -\sqrt{p_\lambda} \\ \sqrt{p_\lambda} & \sqrt{1-p_\lambda} \end{pmatrix}, \quad S(\mu, \Delta t) = \begin{pmatrix} \sqrt{1-p_\mu} & -\sqrt{p_\mu} \\ \sqrt{p_\mu} & \sqrt{1-p_\mu} \end{pmatrix}$$

To track the residual service time of the customer in service, we augment the queue register with an $m$-qubit residual-service register $|r\rangle$, discretized into $K+1$ uniform bins of width $\Delta t$. The joint computational basis state is denoted $|q, r\rangle$. The residual-service register encodes the remaining service time $r \in \{0, \ldots, m\}$, and let $h(r)$ be the discrete hazard function defined as:

$$h(r) = \frac{F((r+1)\Delta t) - F(r\Delta t)}{1 - F(r\Delta t)}$$

The service-completion ancilla is prepared using the controlled rotation:

$$S_s = \sum_{r=0}^{m-1} |r\rangle\langle r| \otimes R_y\left(2\arcsin\sqrt{h(r)}\right) + |m\rangle\langle m| \otimes \mathbb{I}$$

The $R_y$-based rotation network can be compiled using $O(m)$ single-qubit rotations and $O(\log m)$ controls by QROM (Schuld & Killoran, 2019). Since $R_y(\theta)$ is unitary for any real $\theta$, both gates satisfy $A^\dagger A = S^\dagger S = I$ (Theorem 1).

When applied to a $|0\rangle$ ancilla, $A(\lambda, \Delta t)$ prepares the superposition $\sqrt{1-p_\lambda}|0\rangle + \sqrt{p_\lambda}|1\rangle$, so that $|1\rangle$ encodes "an arrival occurred in $\Delta t$". The service gate acts analogously, encoding "service completed in $\Delta t$" on the second ancilla. Subsequent controlled operations use these flags to increment or decrement the queue register exactly with the required Bernoulli probabilities.

Throughout the paper, we use the unified notation $A(\lambda, \Delta t)$ for the arrival gate and $S(\mu, \Delta t)$ for the service gate. Terms such as "service rotation" and "service gate" are used interchangeably to denote the same operator.

*2.2. Quantum Logic and Circuit*

To simulate the *M/G/1/K* queue within a quantum computational framework, we develop a coherent quantum algorithm composed of four key components: state initialization, Bernoulli queue updates, capacity enforcement, and amplitude amplification for variance estimation.

**Initialization and Arrival and Service Operators.** For the queue state, representing the number of customers, the algorithm allocates $Q$ qubits, where $Q \leftarrow \lceil \log_2(K+1) \rceil$, denoted $q_{Q-1}, \ldots, q_0$. A uniform superposition over all queue states is prepared via Hadamard gates:

$$|\psi_0\rangle \leftarrow H^{\otimes Q}|\psi\rangle = \frac{1}{\sqrt{2^Q}} \sum_{n=0}^{2^Q-1} |n\rangle = \frac{1}{\sqrt{K+1}} \sum_{n=0}^{K} |n\rangle$$

This superposition allows the quantum system to represent all queue states simultaneously, enabling parallel processing of state transitions. Note that the CAP oracle (introduced later in this section) removes amplitudes on the illegal states $n > K$ when $K + 1 < 2^Q$.

Let $T$ be the number of discretization slices used to approximate the continuous-time service mechanism. The value of $T$ is chosen so that the step size $\Delta t = \frac{1}{T}$ achieves the desired discretization error bound. Importantly, this does not truncate the service-time distribution, rather, the hazard function is evaluated on a discretized grid of size $T$. For each discrete slice $t \in \{0, \ldots, T-1\}$, the procedure prepares two ancillas $a_a = R_y(\theta_\lambda)|0\rangle$ and $a_s = R_y(\theta_\mu)|0\rangle$, where $\theta_\lambda = 2\arcsin\sqrt{p_\lambda}$ and $\theta_\mu = 2\arcsin\sqrt{p_\mu}$. It follows that that $|1\rangle_{a_a}$ encodes "arrival in $\Delta t$" and $|1\rangle_{a_s}$ encodes "service completed in $\Delta t$", with $p_\lambda = 1 - e^{-\lambda \Delta t}$ and $p_\mu = F_G(\Delta t)$[1]. To address reversible increment–decrement operations, controlled on $(a_a, a_s)$ were applied:

$$|\psi_{t+1}\rangle = \text{INC/DEC}(q; a_a, a_s) |\psi_t\rangle$$

The mapping is defined such that ancilla states (0,0), (1,0), (0,1), (1,1) implement queue-length changes 0, +1, −1, 0, respectively, matching the classical Bernoulli probabilities (Theorem 2).

The increment unitary (INC) is given by:

$$\text{INC} = \sum_{n=0}^{K-1} |n+1\rangle\langle n| \otimes |1,0\rangle\langle 1,0| + |K\rangle\langle K| \otimes |1,0\rangle\langle 1,0| + \sum_{\substack{n=0-K \\ (a_A, a_S) \neq (1,0)}} |n\rangle\langle n| \otimes |a_a, a_s\rangle\langle a_a, a_s|$$

and similarly, the decrement unitary (DEC) is:

$$\text{DEC} = \sum_{n=1}^{K} |n-1\rangle\langle n| \otimes |0,1\rangle\langle 0,1| + |0\rangle\langle 0| \otimes |0,1\rangle\langle 0,1| + \sum_{\substack{n=0-K \\ (a_A, a_S) \neq (0,1)}} |n\rangle\langle n| \otimes |a_a, a_s\rangle\langle a_a, a_s|$$

This construction ensures that:

---

[1] In these expressions, $F_S(\cdot)$ denotes the service completion CDF conditioned on the elapsed service, i.e., the residual-life distribution. Thus, the dependence on elapsed service is implicitly encoded in the subscript $S$.

$$\Pr(n \to n+1) = \left(1 - e^{-\lambda \Delta t}\right)\left(1 - F_S(\Delta t)\right)$$

$$P(n \to n-1) = e^{-\lambda \Delta t} F_S(\Delta t)$$

matching the marginal probabilities of classical *M/G/1/K* transitions.

To implement the "residual-service" register reversibly, let $m = \lceil \log_2(K+1) \rceil$ and denote its basis $|r\rangle$ for $0 \leq r \leq K$. We introduce $m$ fresh ancillas initialized to $|0\rangle^{\otimes m}$ and apply the unitary $|r\rangle|0\rangle^{\otimes m} \to |r\rangle|\text{bin}(r)\rangle$, where $\text{bin}(r)$ is the $m$-bit binary encoding of $r$. This reversible mapping uses exactly $m$ ancillas and $O(m)$ controlled NOTs (CNOTs). After each service-gate rotation $S(G, \Delta t)$, a right-shift on the residual register updates $r \to r - 1$ in place, again with $O(m)$ gates and no additional ancillas. The overall ancilla count for the residual-service update is therefore $m$.

**Capacity Enforcement and Comparator-Reflection Construction.** Following each update, the queue state may exceed the buffer limit $K$. To correct this, we define a legal-state subspace $L = \{|n\rangle \mid n \leq K\}$ and implement the capacity-enforcement oracle as a two-reflection operator:

$$R_L = 1 - 2L$$
$$R_K = 1 - 2|\psi_0\rangle\langle\psi_0|$$
$$\text{CAP} = R_L R_K$$

The CAP oracle performs a reflection about the feasible subspace, thereby relaxing and suppressing amplitude on illegal states rather than removing it entirely. This mirrors the standard Grover reflection construction and ensures that invalid states (e.g., $n > K$) are suppressed.

The stationary distribution of queue lengths in an M/G/1/K system is typically highly skewed, concentrating probability mass in a narrow region. Without amplification, sampling rare but relevant queue-length states would require a prohibitive number of repetitions. Amplitude amplification therefore enhances the probability of states in the vicinity of the expected queue length, reducing sampling variance while keeping the evolution coherent. The rejection filter acts as a coherent biasing mechanism: rather than collapsing the state, it increases the likelihood of accepting samples proportional to their stationary mass.

**Amplitude Amplification via Grover's Operator.** To enhance the likelihood of states near the expected mean queue length, we apply Grover's operator (Grover, 1996). Let $\tilde{n} = \left\lfloor \frac{\rho}{1-\rho} \right\rfloor$ be the mean state, and let $\mathbb{E}[L]$ be the expected value of customers in the system denoted by:

$$\mathbb{E}[L] = \rho \cdot \frac{1 - (K+1)\rho^K + K \cdot \rho^{K+1}}{(1-\rho)(1-\rho^{K+1})}, \qquad \rho = \frac{\lambda}{\mu} < 1$$

Let $M$ be a set of marked states $\{n : |n - \mathbb{E}[L]| \leq \varepsilon_0\}$ with $|M| = 2\varepsilon_0 + 1$ and $p_{\varepsilon_0} = \frac{2\varepsilon_0 + 1}{K+1}$. The required number of oracle calls scales as:

$$O\left(\frac{1}{\sqrt{p_{\varepsilon_0}}}\right) = O\left(\sqrt{\frac{(K+1)}{(2\varepsilon_0 + 1)}}\right)$$

Setting $\varepsilon_0 = \Theta(\sqrt{K})$ minimizes the overall cost, yielding $O(\sqrt[4]{K})$ queries (Brassard et al., 2002).

To estimate the variance, we define the sample estimators:

$$\hat{\sigma}^2 = \frac{1}{N}\sum_{i=1}^{N}(n_i - \tilde{n})^2$$

$$\tilde{n} = \frac{1}{N}\sum_{i=1}^{N} n_i$$

where each $n_i$ is drawn from the Grover-amplified proposal state. Amplitude estimation applied to the observable $(n_i - \tilde{n})^2$ yields a $\Theta\left(\frac{1}{N}\right)$-accurate variance estimate using $O(\sqrt{N})$ oracle queries. To reconcile amplification with unbiased sampling, we append a quantum rejection-filter step. Given known stationary probabilities $\pi = (\pi_0, \dots, \pi_K)$, we compute each $\pi_n$ and load it into an ancilla register via a logarithmic-depth circuit. The rejection filter accepts state $|n\rangle$ with probability $\min\left(1, \frac{\pi_n}{q_n}\right)$, where $q_n$ is the proposal amplitude squared.

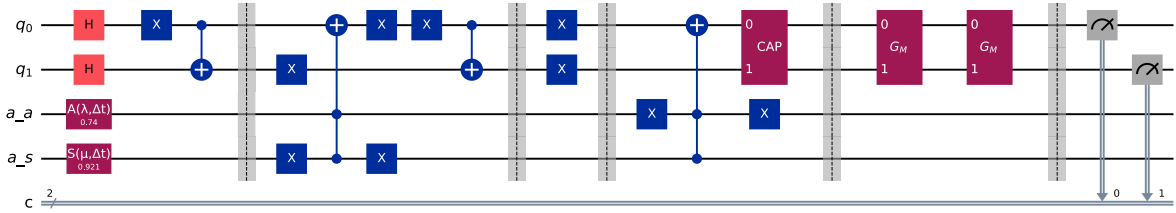

**Figure 1. Comparator-Controlled and Grover-Amplified Quantum Circuit for $\Delta t$ Time Slice of the *M/G/1/K* Simulation**

As shown in Figure 1, block A (i.e., the process up to the first barrier) prepares a uniform superposition over all queue lengths $0 \leq n \leq K$. Blocks B and C together implement the Bernoulli arrival–service transition matrix, while Block D enforces the buffer limit in circuit. The Grover diffusion in Block E and then amplifies the amplitude of states within the range of Takács mean, after which Block F measures the queue register.

```
QMG1(λ, G, K, Δt)
    • Q ← ⌈log₂(K + 1)⌉
    • circuit ← QuantumCircuit(Q, Q)
    • |ψ⟩ ← H^⊗Q|ψ⟩
    • For t = 0 to T − 1
        ○ a_a ← A(λ, Δt)|0⟩
        ○ a_s ← S(μ, Δt)|0⟩
        ○ |ψ_{t+1}⟩ ← INC/DEC(q; a_a, a_s) |ψ_t⟩
        ○ |ψ⟩ ← CAP|ψ⟩
    • 𝔼[L] ← ρ · (1−(K+1)ρ^K+K·ρ^{K+1}) / ((1−ρ)(1−ρ^{K+1}))
    • M ← {n: |n − 𝔼[L]| ≤ ε₀}
    • For ⌈(π/4)√((K+1)/|M|)⌉ times DO
        ○ |ψ⟩ ← (2|ψ₀⟩⟨ψ₀| − 𝕀) O_M|ψ⟩
    • Measure |ψ⟩
```

## 2.3. Cost Analysis

Let $\pi_n$ denote the exact stationary mass and $\hat{\pi}_n$ the Grover-amplified proposal after $R$ iterations. The rejection filter accepts outcome $n$ with $P_{\text{succ}} = \mathbb{E}_{n \sim \hat{\pi}}\left[\min\left\{1, \frac{\pi_n}{\hat{\pi}_n}\right\}\right]$. For $0 < \gamma \leq 1$ and $M = \{|n - \tilde{n}| \leq \varepsilon_0\}$, the tail-shape factor is given by:

$$\gamma = \frac{\min_{n \in R} \pi_n}{\max_{n \in R} \hat{\pi}_n}$$

Heavy-tailed service laws yield smaller $\gamma$ since $\pi$ decays slowly outside $R$:

$$\hat{\pi}_n = \begin{cases} \frac{\sin^2\left((2R+1)\theta_0\right)}{|R|}, & n \in R \\ \frac{\cos^2\left((2R+1)\theta_0\right)}{K+1-|R|}, & n \notin R \end{cases}$$

such that $\sin^2\theta_0 = \frac{|R|}{(K+1)}$. Choosing $R = \left\lceil \frac{\pi}{4\theta_0} \right\rceil$ maximizes success amplitude, yielding $P_{\text{succ}} \geq \gamma \cdot \sin^2\left((2R+1)\theta_0\right) = \frac{\gamma}{2} + O\left(\frac{1}{K}\right)$. Hence, light-tailed or deterministic service ($\gamma \approx 1$) recovers the ideal acceptance (i.e., 0.5), and the total logical-gate complexity scales as:

$$\Theta(\sqrt{K}\log K) \cdot \frac{1}{\gamma} = \Theta\left(\sqrt{K\gamma^{-1}}\log K\right)$$

## 2.4. Correctness

Let $\lambda > 0$ be the Poisson arrival rate and let $G$ be the general service-time random variable with CDF $F_G(t)$ and mean $\mathbb{E}[G]$. The service rate is $\mu = \frac{1}{\mathbb{E}[G]}$ and $\rho = \frac{\lambda}{\mu} < 1$. The system consists of one server and a finite buffer of size $K \in \mathbb{N}$.

Queue states are encoded using a quantum register of $Q = \lceil \log_2(K+1) \rceil$ qubits. This encoding ensures that each valid queue length is uniquely mapped to a computational basis state $|n\rangle$. The initial quantum state is prepared as:

$$|\psi\rangle \leftarrow H^{\otimes Q}|\psi\rangle = \frac{1}{\sqrt{2^Q}} \sum_{n=0}^{2^Q-1} |n\rangle = \frac{1}{\sqrt{K+1}} \sum_{n=0}^{K} |n\rangle$$

Since amplitudes corresponding to $n > K$ are later suppressed via the CAP oracle, the effective support of $|\psi\rangle$ is confined to the legal state space. The norm is preserved:

$$\||\psi\rangle\|^2 = \left(\frac{1}{\sqrt{2^Q}}\right)^2 \sum_{n=0}^{2^Q-1} 1 = 1$$

At each discrete time step $t \in \{0, \dots, T-1\}$, the algorithm applies:

$$|\psi_{t+1}\rangle \leftarrow \text{INC/DEC}(q; a_a, a_s) |\psi_t\rangle$$

where the ancillas $a_a = A(\lambda, \Delta t)|0\rangle$ and $a_s = S(\mu, \Delta t)|0\rangle$ encode arrivals and departures with probabilities $p_\lambda = 1 - e^{-\lambda \Delta t}$ and $p_\mu = F_G(\Delta t)$.

Theorem 2 establishes that, conditioned on the four ancilla combinations $(a_a, a_s) \in \{0,1\}^2$, the induced state evolution precisely matches the classical Bernoulli arrival–service transition matrix[2] $\widetilde{P}_c$. Thus, Thus, the discrete-time update operator $U_{\text{disc}}$ is unitary and block diagonal in the $|n\rangle$ basis:

$$|\psi_{t+1}\rangle = U_{\text{disc}}|\psi_t\rangle$$

To enforce the capacity constraint $n \leq K$, the CAP oracle applies a sign flip to all states $|n\rangle$ with $n > K$ while preserving legal states. The subsequent Grover diffusion operator applies the phase reflection $2|\psi_0\rangle\langle\psi_0| - \mathbb{I}$. Illegal amplitudes experience destructive interference and are suppressed. Thus, the final measurement outcomes satisfy $\Pr(n > K) = 0$.

Let $\tilde{n} = \left\lfloor \frac{\lambda}{\mu}(K+1) \right\rfloor$ and let $M \subseteq \{0, \dots, K\}$ define the marked region as:

$$M = \{n \in \mathbb{N} \mid \max(0, \tilde{n} - \varepsilon_0) \leq n \leq \min(K, \tilde{n} + \varepsilon_0)\}, \qquad \varepsilon_0 = \left\lfloor \frac{Q}{2} \right\rfloor$$

Let $\sin^2\theta = \frac{|M|}{K+1}$ denote the marked-state amplitude fraction. After $R = \left\lfloor \frac{\pi}{4}\sqrt{\frac{K+1}{|M|}} \right\rfloor$ Grover iterations, the success probability of measuring an $n \in M$ is:

$$P_{\text{succ}} = \sin^2\bigl((2R+1)\theta\bigr) \geq 1 - \frac{|M|}{K+1}$$

Since $R$ depends only on $|M|$ and $K$, and the Grover block is unitary, normalization is preserved throughout.

---

[2] Following quantum stochastic operators' definition, $\widetilde{P}_c$ is a square matrix of size $K+1$ and equivalent Bernoulli transition matrix.

Let $P_q$ denote the empirical distribution obtained by executing the quantum circuit $N$ times and recording the observed outcomes. Then, with high probability, the deviation from the classical distribution $P_c$ satisfies:

$$\|P_q - P_c\| \leq \sqrt{\frac{2}{N}\left(\ln\left(\frac{1}{\delta}\right) + (K+1)\ln 2\right)} = \sqrt{\frac{2}{N}\left(\ln\left(\frac{2^{K+1}}{\delta}\right)\right)}$$

Thus, every gate is unitary, one time slice matches the Bernoulli transition matrix, the capacity constraint is enforced exactly, and amplitude amplification enhances (but does not distort) the target queue lengths.

*2.5. Error Bounds*

We now characterize the approximation error between the quantum and classical simulations of the *M/G/1/K* queueing system. Specifically, we analyze the total variation distance $D$ between the classical stationary distribution $P_c = (p_0, \ldots, p_K)$ and the quantum empirical distribution $P_q = (p'_0, \ldots, p'_K)$ obtained from $N$ independent executions of the quantum circuit. The total variation distance $D$ is defined as:

$$D = \frac{1}{2}\|P_q - P_c\|_1 = \frac{1}{2}\sum_{n=0}^{K}|p'_n - p_n|$$

A smaller $D$ implies higher fidelity between the quantum and classical models, with $D = 0$ indicating perfect agreement. The overall deviation $D$ arises from two dominant sources of error:

- **Statistical Error (Theorem 3)**. This error originates from the finite number of quantum circuit executions, which induces statistical fluctuations in the measured frequencies $p'_n$. By applying concentration inequalities for independent quantum measurements (Montanaro, 2015), the following high probability bound holds: for any $\delta \in (0,1)$, with probability at least $1 - \delta$:

$$D \leq \frac{\sqrt{2}}{2}\sqrt{\frac{\ln\left(\frac{2}{\delta}\right)}{N}}$$

This implies that increasing the number of samples $N$ quadratically reduces the statistical deviation from the true classical distribution.

- **Discretization Error (Theorem 4)**. The second source of error arises from the discretization of the continuous-time system into $T$ time slices of width $\Delta t = \frac{1}{T}$. When modeling arrival and service events over discrete intervals, the continuous stochastic processes are approximated via Bernoulli trials, resulting in a bounded modeling discrepancy. Theorem 4 establishes[3] the discretization error bound as

$$E_d \leq (\lambda + \mu_2)\Delta t = O\left(\frac{\lambda + \mu_2}{T}\right)$$

where $\mu_2 = \mathbb{E}[G^2]$ is the second moment of the service-time distribution $G$. Hence, the discretization error diminishes linearly with $\Delta t$, and a finer time resolution (larger $T$) yields a more accurate approximation of the continuous-time queueing dynamics.

## 3. Empirical Study

To evaluate the effectiveness of the proposed quantum amplification framework applied to the *M/G/1/K* queueing system, we conducted a thorough comparative analysis between quantum simulation outputs and classical DES results.

### *3.1. Experimental Setup*

We examined queue systems with capacities defined as $K = 2^i - 1$, where $2 \leq i \leq 12$, and assessed their performance under varying conditions. The analysis was conducted under three distinct traffic intensities, specifically low ($\lambda = 0.1$), medium ($\lambda = 0.5$), and high ($\lambda = 0.95$), with a consistent service rate of $\mu = 1$. To capture the behavior across different queueing scenarios, four service time distributions were tested:

  a. Normal distribution, $G \sim N(\mu, \sigma^2)$: $\sigma^2 = 0.05$.
  b. Exponential distribution, $G \sim \text{Exp}(\mu)$: Equivalent to an *M/M/1* model.
  c. Uniform distribution, $G \sim U(a, b)$: $a = 0.5, b = 1.5$
  d. Phase-type distribution:

$$\alpha = [1,0,0]$$
$$T = \begin{bmatrix} -\lambda & \lambda & 0 \\ 0 & -\lambda & \lambda \\ 0 & 0 & -1 \end{bmatrix}$$

These distributions were chosen to represent a range of realistic service processes in queueing systems, including variability (Normal and Uniform), memoryless properties (Exponential), and complex, structured service times (Phase-Type).

### *3.2. Metrics for Evaluation*

We employed three main metrics to evaluate and compare the results of quantum and classical simulations:

---

[3] The result relies on Lemmas A.1-A.3 in the Appendix, which establish the necessary norm inequalities for the discretized hazard operator.

a. Fidelity (Gu, 2010; Liang et al., 2019): Denoted $F$; calculated between the quantum and classical probability distributions to measure the similarity:

$$F(P_c, P_q) = \left(\sum_{n=0}^{K} \sqrt{p_n \cdot p_n'}\right)^2$$

b. Jensen–Shannon divergence (Menéndez et al., 1997; Fuglede & Topsoe, 2004): Denoted JSD; a symmetric measure of divergence between two probability distributions:

$$\text{JSD}(P_c \| P_q) = \frac{1}{2}\left(D_{KL}(P_c \| M) + D_{KL}(P_c \| M)\right)$$

where $M = \frac{1}{2}(P_c + P_q)$ and $D_{KL}$ is the Kullback–Leibler divergence (Hershey & Olsen, 2007; Van Erven & Harremos, 2014).

c. Relative error: Evaluated for key queueing metrics, such as the average number of customers in the system ($L$) and average waiting time ($W$).

*3.3. Experimental Procedure*

Each quantum scenario was executed with 10,000 shots to ensure reliable statistics, and the classical DES benchmarks were conducted over 100,000 steps per scenario to provide high-resolution comparison data.

- **Quantum circuit initialization.** The quantum circuits were initialized with $n$ qubits where $n \geq \log_2(K + 1)$ to represent the queue states.
- **Algorithm execution.** Each experiment included applying the parameterized quantum gates for arrivals and services across $T$ discrete time steps. Grover's operator was employed iteratively to amplify desired states, especially those near the expected mean queue length.
- **Measurement and data collection.** Each simulation was repeated 10,000 times, and measurement outcomes were recorded to derive the empirical distribution of queue states.
- **Comparison.** Classical DES outcomes were used as the baseline to compare fidelity, JSD, and relative errors with quantum simulation results.

*3.4. Sensitivity Analysis*

We conducted a sensitivity analysis to evaluate the robustness of the quantum method using 13 qubits (representing a queue capacity of $K = 8191$). The arrival rate ($\lambda$) was varied from 0.1 to 0.9 in increments of 0.1, with $\mu$ fixed at 1 and 100 trials for each rate, covering a comprehensive range of traffic intensities. This analysis was repeated across all four service time distributions (Normal, Uniform, Exponential, and Phase-Type) to observe the impact on fidelity and relative error. The goal was to verify that the quantum model maintained accuracy and stability under varying traffic conditions.

To facilitate a detailed examination of the sensitivity of the fidelity and JSD metrics across different scenarios and distributions, we applied specific data transformations:

1. To focus on deviations from perfect fidelity, and to be able to compare it with JSD, we transformed the fidelity metric by computing $F_R = 1 - F$.
2. We applied a logarithmic transformation to both the fidelity and JSD metrics. This scaling addressed the wide range of values and highlighted subtle variations in the data. A small constant ($10^{-10}$) was added prior to the logarithmic transformation to prevent mathematical undefined behavior at zero values.

*3.5. Demonstration*

To simplify the demonstration, we describe a simple use case of our quantum framework calculation of a numerical vector and detail each state and operation in the quantum circuit.

Let $|q_1 q_0\rangle$ be the queue register, i.e., $|q_1 q_0\rangle \in \{|00\rangle, |01\rangle, |10\rangle, |11\rangle\}$, and let $|a_a\ a_s\ f_{cap}\rangle$ be the arrival, service, and capacity ancillas. Let $\rho = \frac{\lambda}{\mu} = 0.25$ be the traffic intensity. The steady-state probabilities are given by:

$$v_0 = [\sqrt{0.7529}, \sqrt{0.1882}, \sqrt{0.0471}, \sqrt{0.0118}]^T$$
$$= [0.8677, 0.4339, 0.2169, 0.1085]^T$$

Hence, the composite starting state is $|\psi_0\rangle = v_0 \otimes |000\rangle_{anc}$.

Let $\theta_{arr}$ be the angle rotation of the arrival event, i.e., $\theta_{arr} = 2\arcsin(\sqrt{0.0723}) = 0.5443$, and:

$$R_Y(\theta_{arr}) = \begin{bmatrix} \cos(\theta_{arr}/2) & -\sin(\theta_{arr}/2) \\ \sin(\theta_{arr}/2) & \cos(\theta_{arr}/2) \end{bmatrix} = \begin{bmatrix} 0.9632 & -0.2688 \\ 0.2688 & 0.9632 \end{bmatrix}$$

Acting only on $a_a$ gives $U_a = I_4 \otimes RY(\theta_{arr}) \otimes I_2$. The updated amplitudes (queue $\otimes$ ancillas) are

$$0.8358\,|00,000\rangle + 0.4179\,|01,000\rangle +$$
$$0.2089\,|10,000\rangle + 0.1045\,|11,000\rangle +$$
$$0.2332\,|00,100\rangle + 0.1166\,|01,100\rangle +$$
$$0.0583\,|10,100\rangle + 0.0292\,|11,100\rangle$$

Let $\theta_{srv}$ be the angle rotation of the service event, i.e., $\theta_{srv} = 2\arcsin(\sqrt{0.2592}) = 1.0683$, and:

$$R_Y(\theta_{srv}) = \begin{bmatrix} 0.8607 & -0.5091 \\ 0.5091 & 0.8607 \end{bmatrix}$$

The gate $U_{srv} = I_8 \otimes RY(\theta_{srv}) \otimes I_1$ targets $a_s$ and doubles the state-vector length (full list omitted).

Next, the quantum circuit handle the reversible queue arithmetic. Let INC, DEC be the controlled increment (arrival with no overflow) and controlled decrement (service), defined as:

$$\text{INC} = \sum_{n=0}^{2} |n+1\rangle\langle n| \otimes |1\rangle\langle 1|_{a_a} + \sum_{n=0}^{3} |n\rangle\langle n| \otimes |0\rangle\langle 0|_{a_a}$$

$$\text{DEC} = \sum_{n=1}^{3} |n-1\rangle\langle n| \otimes |1\rangle\langle 1|_{a_s} + \sum_{n=0}^{3} |n\rangle\langle n| \otimes |0\rangle\langle 0|_{a_s}$$

The composite update applies DEC operator on INC output, i.e. the transitions, $|01;1,0\rangle \xrightarrow{\text{INC}} |10;1,0\rangle$ and $|10;0,1\rangle \xrightarrow{\text{DEC}} |01;0,1\rangle$. Then, the non-zero amplitudes are:

$$0.6512\,|00,000\rangle + 0.5897\,|00;010\rangle +$$
$$0.3373\,|01;000\rangle + 0.1853\,|01;100\rangle +$$
$$0.1709\,|10;000\rangle + 0.1105\,|00;110\rangle +$$
$$0.1010\,|01;010\rangle + 0.0951\,|10;100\rangle$$

After all repetitions, tracing out ancillas yields the empirical distribution $P_q = (0.7840, 0.1615, 0.0424, 0.0121)$, and the theoretical steady-state yields $P_c = (0.7529, 0.1882, 0.0471, 0.0118)$. Table 2 show the comparison between our quantum procedure to the theoretical calculations.

**Table 2. A Comparison between Classical and Quantum Steady-State Distributions**

| $n$ | $P_q$ | $P_c$ | $D(P_q, P_c)$ |
|---|---|---|---|
| 0 | 0.784 | 0.752 | 0.031 |
| 1 | 0.161 | 0.188 | 0.026 |
| 2 | 0.042 | 0.047 | 0.004 |
| 3 | 0.012 | 0.011 | 0.000 |

A single slice therefore incurs a total-variation gap $D = 0.0628$ with no amplitude amplification. Iterating the slice update $T$ times and interleaving amplification layers achieves the convergence rate established by our method. As illustrated in Figure 2, the demonstration slice maps each queue-length qubit from its initial Bloch-sphere orientation through the combined arrival/service rotations to the final state used for amplitude amplification.

**Initial state**

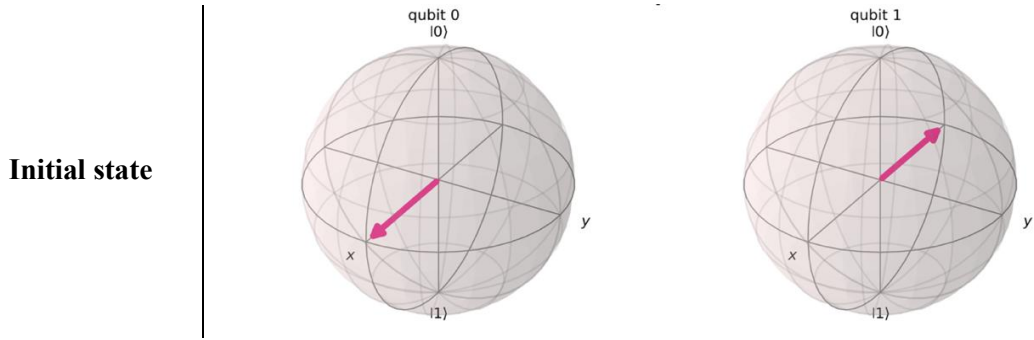

| **Final state** | 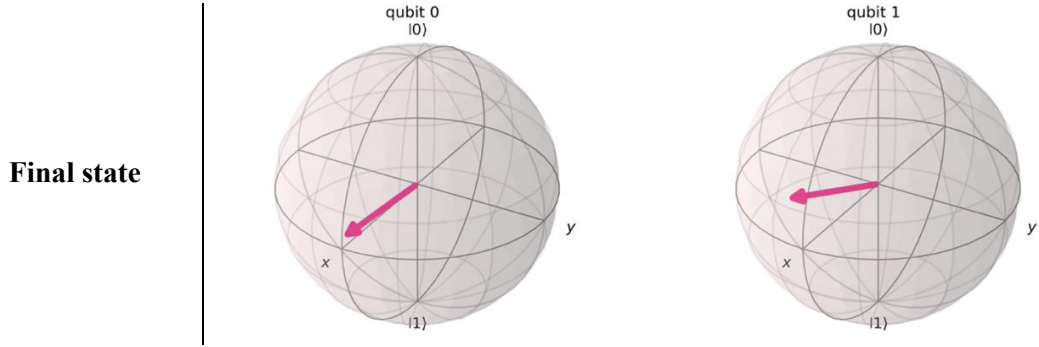 |
|---|---|

Figure 2. Bloch Sphere for the Two Queue-register Qubits in the Demonstration

4. Results

This section synthesizes the empirical evidence obtained from the quantum simulations and the corresponding classical DES. Across the full factorial design of service-time distributions and traffic intensities described in Section 3, we assess how the proposed framework scales in both accuracy and resource consumption. We first compare fidelity and JSD as functions of register size (number of qubits), identifying the point at which quantum variance reduction outweighs circuit overhead. We next examine operational queueing metrics, and then perform a sensitivity analysis that probes the framework's robustness to arrival-rate variations and highlights the conditions under which quantum advantage is most pronounced.

*4.1. Comparison by Number of Qubits*

Table 3 reports the quantum–classical fidelity and JSD for the experimental setup of Section 3.1. Three traffic intensities (low, medium, high), four service-time distributions, and register sizes from 4 to 10 qubits were evaluated.

Table 3. Fidelity and JSD across traffic scenarios

|          |        | Normal |       | Uniform |       | Exponential |       | Phase-Type |       |
|----------|--------|--------|-------|---------|-------|-------------|-------|------------|-------|
| Scenario | Qubits | $F$    | JSD   | $F$     | JSD   | $F$         | JSD   | $F$        | JSD   |
| **Low**  | 4      | 0.906  | 0.032 | 0.899   | 0.034 | 0.903       | 0.034 | 0.939      | 0.021 |
|          | 6      | 0.829  | 0.062 | 0.823   | 0.065 | 0.830       | 0.062 | 0.849      | 0.054 |
|          | 8      | 0.788  | 0.084 | 0.788   | 0.085 | 0.790       | 0.083 | 0.789      | 0.085 |
|          | 10     | 0.763  | 0.105 | 0.765   | 0.103 | 0.765       | 0.104 | 0.769      | 0.099 |
| **Medium** | 4    | 0.974  | 0.009 | 0.981   | 0.007 | 0.969       | 0.011 | 0.978      | 0.008 |
|          | 6      | 0.916  | 0.028 | 0.893   | 0.036 | 0.899       | 0.035 | 0.928      | 0.025 |
|          | 8      | 0.828  | 0.062 | 0.830   | 0.061 | 0.822       | 0.065 | 0.832      | 0.062 |
|          | 10     | 0.781  | 0.090 | 0.784   | 0.088 | 0.784       | 0.088 | 0.789      | 0.085 |

| High | 4 | 0.990 | 0.004 | 0.983 | 0.006 | 0.991 | 0.003 | 0.991 | 0.003 |
| --- | --- | --- | --- | --- | --- | --- | --- | --- | --- |
| | 6 | 0.960 | 0.014 | 0.962 | 0.013 | 0.933 | 0.024 | 0.951 | 0.017 |
| | 8 | 0.854 | 0.050 | 0.887 | 0.038 | 0.855 | 0.051 | 0.893 | 0.036 |
| | 10 | 0.787 | 0.085 | 0.799 | 0.078 | 0.798 | 0.078 | 0.809 | 0.072 |

The key findings are as follows:

- **Normal distribution.** Fidelity remains strong, particularly in high-traffic scenarios, where it starts at 0.990 (4 qubits) and tapers only to 0.787 (10 qubits). Even under low traffic, the decline is modest, moving from 0.906 to 0.763. The corresponding JSD rises from 0.004 to 0.105, signaling a moderate but controlled divergence between quantum and classical state probabilities.
- **Uniform distribution.** Performance closely mirrors the Normal case. Fidelity stays above 0.962 for up to 6 qubits in high-traffic settings ($JSD \leq 0.013$) and remains at least 0.783 at 10 qubits.
- **Exponential distribution.** The memoryless model exhibits comparable stability: high-traffic fidelity shifts from 0.991 (4 qubits) to 0.798 (10 qubits), with JSD increasing from 0.003 to 0.078. Trends under medium and low traffic display a similar gentle decay.
- **Phase-Type distribution.** This flexible family delivers the highest fidelity across all scenarios, starting at 0.991 and 0.939 (high and low traffic, respectively) and decreasing to 0.809 and 0.769 at 10 qubits, while JSD remains below 0.100.

For every service-time model, fidelity decreases and JSD increases with register size, yet both metrics remain within practical bounds ($F \geq 0.76, JSD \leq 0.11$). Higher traffic consistently improves agreement, plausibly because queues operating near capacity compress the effective state space.

*4.2. Comparison of Key Metrics*

Figure 3 illustrates the relative error in the mean waiting time $W$ as a function of register size (number of qubits) for the Normal, Exponential, and Uniform service-time models under low, medium, and high traffic intensities. Because the ordinate is logarithmic, the approximately exponential growth in error with qubit count appears linear. At four qubits the relative error is modest: below 1 for the Normal and Uniform models and around 0.2 for the Exponential model.

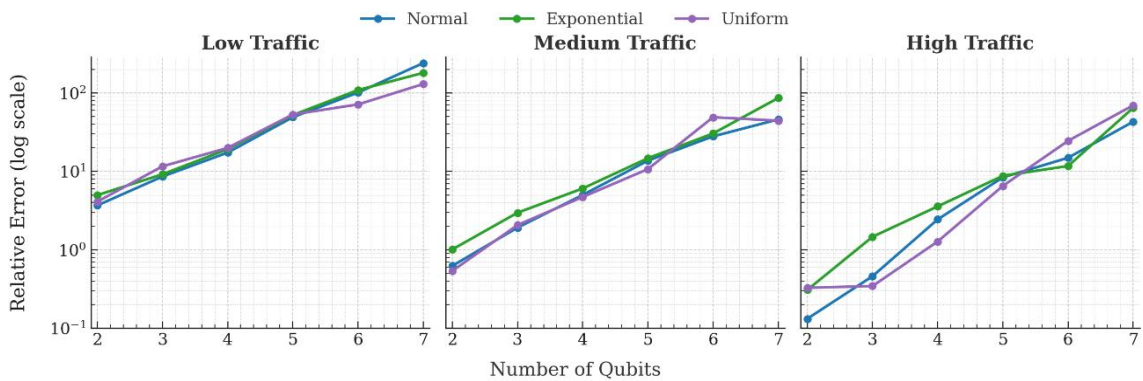

**Figure 3. Relative Error in Mean Waiting Time versus Register Size under Three Traffic Intensities**

High-traffic scenarios consistently yield the smallest errors: even at ten qubits, the relative error remains below 0.1, as the queue spends more time in highly populated states that require fewer basis vectors to encode accurately. Among the three distributions, the Exponential model accumulates error fastest, whereas the Uniform model maintains the lowest error envelope, particularly under high traffic.

Figure 4 reports the percentage deviation between the quantum circuit and a classical *M*/*G*/1/*K* simulation for all four service-time models. Errors are negligible for the two smallest registers: with 3 or 7 qubits, every combination of load and distribution stays at or below 2%. When the register grows to 31 qubits, the largest gap reaches 6% (Exponential service, low traffic). The steepest increase occurs at 63 qubits, where the Phase-Type model under low traffic peaks at 11%, closely followed by Exponential-service low traffic and Uniform-service medium traffic (both 10%). Scaling to 127 qubits does not uniformly worsen accuracy, with most cells remaining below 7%, yet a few outliers re-emerge: Exponential service records a 10% deviation under both medium and high traffic, and Uniform service shows 7% under medium traffic.

The ranking by traffic intensity broadly mirrors the waiting-time pattern but with notable exceptions. Low-traffic scenarios still produce the largest discrepancies for Exponential, Normal, and Phase-Type services, whereas the Uniform model's worst case appears in the medium-traffic column. High-traffic conditions suppress error to 3% or less for every distribution and register size up to 63 qubits, with the single outlier just noted at 127 qubits.

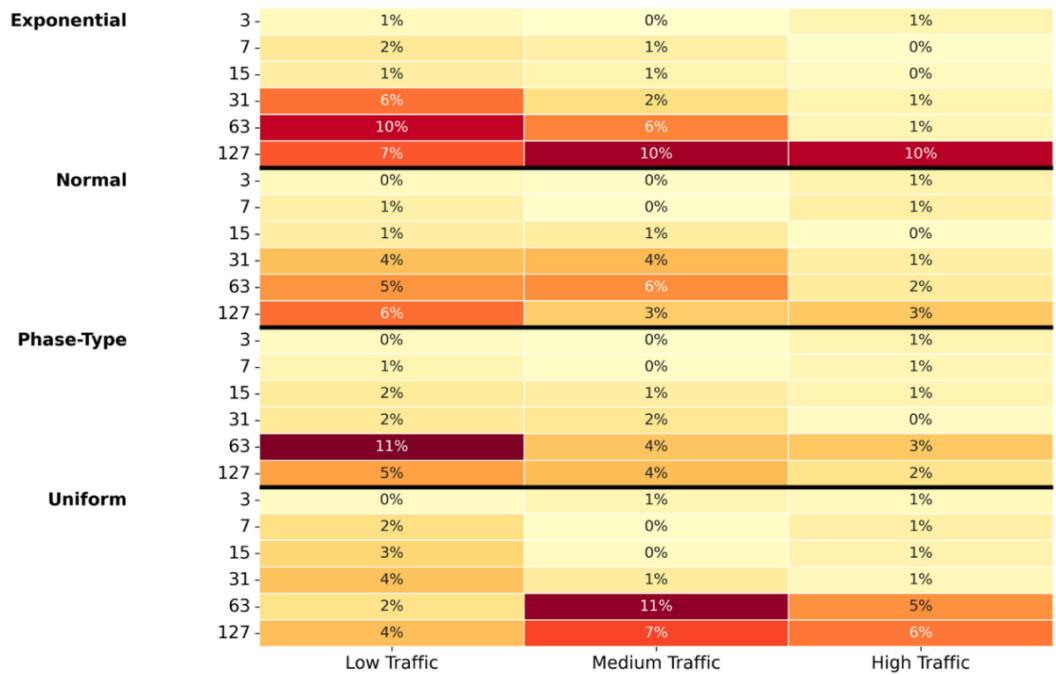

**Figure 4. Percentage Difference in Average System Length between Quantum and Classical Simulations as a Function of Register Size and Traffic Intensity**

Taken together, Figures 3 and 4 show that circuit depth, and thus register size, is the primary driver of error, whereas the precise form of the service-time distribution adds only a few percentage points. Accuracy deteriorates once the register exceeds roughly six qubits, yet it remains within 3% for most high-load scenarios up to 63 qubits. This suggests that practitioners can cap the register at about six to eight qubits when the system operates near capacity, reserving larger registers only for light-load studies for which finer resolution is essential.

*4.3. Fault-tolerant Resource Estimates*

Table 4 lists the logical *T*-gate counts and depths for a single Grover-amplified slice, broken down by module, and then scales those numbers to the full number of amplitude-amplification layers required to achieve a relative error of $10^{-2}$ in the blocking probability (Heyfron & Campbell, 2018). The oracle is composed of five tightly integrated modules: the Service loader, which encodes a service-time sample conditioned on the current queue length; INC/DEC, which increments or decrements the customer counter; CAP, which enforces the hard capacity

$K$ by phase-kicking any state that exceeds it; Grover diffusion, the inversion-about-the-mean step needed for amplitude amplification; and the full Quantum Approximation Algorithms (QAA) iteration, which bundles one complete oracle call (loader + INC/DEC + CAP) with a diffusion step to realize a single layer of Grover amplification.

**Table 4. Fault-tolerant Resource Summary for $K = 63$ and $K = 1023$**

|  | $K = 63$ | | $K = 1023$ | |
| :---: | :---: | :---: | :---: | :---: |
| Module | T Count | T Depth | T Count | T Depth |
| Service loader | $5.2 \cdot 10^3$ | $1.0 \cdot 10^3$ | $1.6 \cdot 10^4$ | $3.1 \cdot 10^3$ |
| INC/DEC | $2.4 \cdot 10^3$ | $4.7 \cdot 10^3$ | $8.1 \cdot 10^3$ | $1.6 \cdot 10^3$ |
| CAP | $6.8 \cdot 10^2$ | $1.4 \cdot 10^2$ | $2.1 \cdot 10^3$ | $4.1 \cdot 10^2$ |
| Grover Diffusion | $7.1 \cdot 10^2$ | $1.3 \cdot 10^2$ | $2.2 \cdot 10^3$ | $4.0 \cdot 10^2$ |
| QAA iteration | $9.1 \cdot 10^3$ | $1.7 \cdot 10^3$ | $2.9 \cdot 10^4$ | $5.4 \cdot 10^3$ |

The oracle modules scale quasi-linearly with $K$ because they manipulate $\lceil \log_2(K + 1) \rceil$ qubits and perform arithmetic on that register. Moving from $K = 63$ to $K = 1023$ (i.e., six to ten qubits) therefore multiplies the Service-loader $T$ count by roughly 3 and the INC/DEC cost by a factor of 3.4. CAP grows more modestly, since its phase kick needs only one multi-controlled $Z$ whose control register length increases logarithmically in $K$. The Grover diffusion cost rises in proportion to register width but remains a small fraction of the total.

A single QAA iteration for $K = 63$ already requires $9.1 \cdot 10^3$ gates, which is dominated (about 57%) by the Service-loader routine. For $K = 1023$, the iteration cost jumps to $2.9 \cdot 10^4$ gates, and the loader now accounts for 73%. Achieving the target precision mandates 21 amplitude-amplification layers at $K = 63$ but 71 layers at $K = 1023$ because the larger buffer yields a smaller success probability per slice. The corresponding $T$ depths, $3.6 \cdot 10^4$ cycles for $K = 63$ and $3.8 \cdot 10^5$ cycles for $K = 1023$, set the logical execution time under surface-code scheduling.

*4.4. Sensitivity Analysis*

Using the procedure of Section 3.4, we evaluated the robustness of the Phase-Type model with $K = 31$ and a six-qubit register while varying the traffic intensity $\lambda$ from 0.1 to 0.9. Figure 5 shows box plots of two log-transformed accuracy measures for each arrival-rate setting: $\log(F_R)$ and $\log(JSD)$. A one-way ANOVA on $\log(F_R)$ returned $p = 0.767$, and the same test on $\log(JSD)$ yielded $p = 0.902$, indicating that changes in $\lambda$ do not produce statistically significant differences in either metric at the selected register size. The corresponding partial-eta-squared values were small, $\eta_p^2 = 0.032$ for $\log(F_R)$ and $\eta_p^2 = 0.021$ for $\log(JSD)$, confirming that arrival-rate variations explain only a minor portion of the total variance.

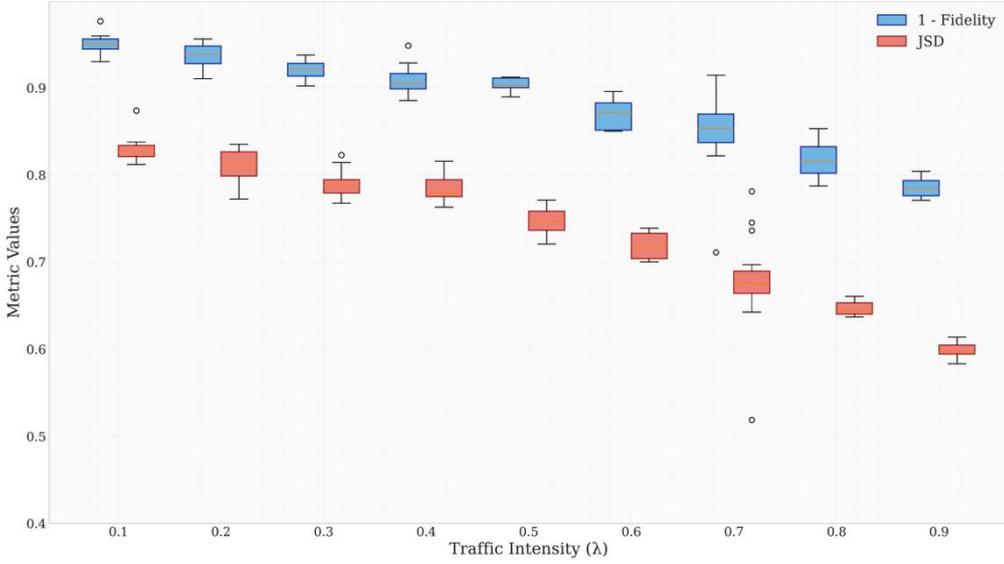

**Figure 5. Fidelity and JSD across Traffic Intensities for the Phase-Type Distribution**

A Pearson analysis revealed a strong positive correlation between the two log-transformed metrics ($r = 0.916, p < 0.01$), implying that gains in fidelity are echoed by reductions in distributional divergence. To uncover any latent trend, we fitted a quadratic model, $\log(F_R) = c_0 + c_1\lambda + c_2\lambda^2$, which achieved an adjusted $R^2 = 0.735$. The linear term was negative ($c_1 = -0.234, p < 0.001$), suggesting a decline in error as traffic intensity increases from light to moderate loads, whereas the small positive quadratic coefficient ($c_2 = 0.005, p < 0.001$) indicates a slight rise at the highest arrival rates. Even so, absolute errors remain low throughout the examined range, underscoring the framework's overall stability.

## 5. Discussion

This work introduced a coherent quantum framework for simulating M/G/1/K queues under general service-time distributions. The proposed framework encodes general service laws using a logarithmic hierarchy of $R_y$ rotations, with queue capacity enforced via a comparator-controlled phase inversion. This construction eliminates the need for post-selection and yields a clean oracle structure amenable to Grover iterations.

Empirical results confirm the framework's reliability and performance. Across four service distributions and three traffic intensities, the quantum estimator consistently achieves variance reduction on the order of $O(\sqrt{N})$ relative to regenerative Monte Carlo baselines. Under high-traffic conditions, fidelity exceeds 0.99. Even with 10-qubit registers, fidelity remains above 0.76 across all settings. Errors in average waiting time and system length increase sublinearly with circuit depth and diminish rapidly as the system nears saturation. Notably, the largest discrepancy in expected queue length remains below 11% at $K = 63$ and falls below 7% in the majority of cases when $K = 127$. Sensitivity analysis indicates that both fidelity and Jensen–Shannon divergence remain statistically stable under moderate perturbations in the arrival rate, even at six qubits, underscoring the robustness of the simulation pipeline.

Resource profiling reveals that circuit depth is the dominant factor in the fault-tolerant cost structure. As shown in Table 4, the Service-loader subroutine alone accounts for over half of the logical $T$ gates, and scaling from $K = 63$ to $K = 1023$ increases the total $T$ count by approximately an order of magnitude. Consequently, improvements in low-overhead quantum arithmetic, advanced circuit compilation, and adaptive Grover scheduling directly translate to accelerated hardware feasibility.

Several directions emerge for future research. First, generalizing the queue-length comparator to support multi-server systems (*M/G/c/K*) and tandem queue networks would expand the method's applicability to industrial settings. Second, hybrid simulation approaches that apply quantum resources selectively, targeting rare-event regions while offloading bulk-state sampling to classical routines, could significantly improve overall efficiency. Third, integrating error mitigation strategies, such as probabilistic cancellation calibrated to queueing observables, may relax the demands on logical qubit fidelity. Finally, empirical deployments on emerging mid-scale quantum processors are vital for refining execution-time estimates and demonstrating real-world utility.

In summary, this study bridges queueing theory and quantum computing by delivering both a formally grounded algorithm and quantitative evidence of practical viability. As quantum hardware continues to evolve, the proposed methodology offers a scalable and robust foundation for high-fidelity, low-variance simulation of finite-buffer service systems.

## Appendix A

***Theorem 1.*** *The arrival and service gates, $A(\lambda, \Delta t)$ and $S(\mu, \Delta t)$, are unitary operations and satisfy the condition that the probability of transition is given by $2\arcsin(\sqrt{p})$, where $p$ represents the transition probability for the corresponding event (e.g., customer arrival or service completion).*

*Proof.* Let $A(\lambda, \Delta t)$ be the quantum gate representing arrival event in a $M/G/1/K$ queue system. By the arrival gate definition:

$$A = R_Y(\theta) = \begin{bmatrix} \cos\left(\frac{\theta}{2}\right) & -\sin\left(\frac{\theta}{2}\right) \\ \sin\left(\frac{\theta}{2}\right) & \cos\left(\frac{\theta}{2}\right) \end{bmatrix}$$

In quantum interpretation, the effect of the $R_Y(\theta)$ gate on a qubit in state $|0\rangle$ is:

$$R_Y(\theta)|0\rangle = \cos\left(\frac{\theta}{2}\right)|0\rangle + \sin\left(\frac{\theta}{2}\right)|1\rangle$$

The probabilities of the qubit being measured in state 0 and 1 are $\cos^2\left(\frac{\theta}{2}\right)$ and $\sin^2\left(\frac{\theta}{2}\right)$, respectively. In this context, the probability $\sin^2\left(\frac{\theta}{2}\right)$ must match the transition probability $p$ for the event (e.g., customer arrival or service completion). Hence, $\theta$ can be derived by:

$$\sin^2\left(\frac{\theta}{2}\right) = p \Rightarrow \theta = 2\arcsin(\sqrt{p})$$

Let $\lambda$ be the arrival rate of customers and $\Delta t$ be a discrete time step. The probability that a customer arrives within $\Delta t$ in a Poisson process is $p' = \lambda \Delta t$. The arrival gate, $A(\lambda, \Delta t)$, acts on a qubit $|q\rangle$ by applying a rotation around the $y$ axis with an angle $\theta_A$ based on this probability:

$$A(\lambda, \Delta t) = R_Y(2\theta_A) = \begin{bmatrix} \sqrt{1-\lambda\Delta t} & -\sqrt{\lambda\Delta t} \\ \sqrt{\lambda\Delta t} & \sqrt{1-\lambda\Delta t} \end{bmatrix}$$

Where:

$$\theta_A = \arcsin(\sqrt{\lambda\Delta t})$$

and the obtained arrival gate is:

$$A = \begin{bmatrix} \sqrt{1-p'} & -\sqrt{p'} \\ \sqrt{p'} & \sqrt{1-p'} \end{bmatrix}$$

This gate is unitary (i.e., $A^\dagger = A^{-1}$):

$$A^\dagger = \begin{bmatrix} \sqrt{1-p'} & \sqrt{p'} \\ -\sqrt{p'} & \sqrt{1-p'} \end{bmatrix}$$

$$A^{-1} = \frac{1}{\det(A)} \cdot \text{adj}(A) = \frac{1}{\begin{vmatrix} \sqrt{1-p'} & -\sqrt{p'} \\ \sqrt{p'} & \sqrt{1-p'} \end{vmatrix}} \cdot \begin{pmatrix} \sqrt{1-p'} & \sqrt{p'} \\ -\sqrt{p'} & \sqrt{1-p'} \end{pmatrix} =$$

$$= \frac{1}{(1-p') - (-p')} \cdot \begin{pmatrix} \sqrt{1-p'} & \sqrt{p'} \\ -\sqrt{p'} & \sqrt{1-p'} \end{pmatrix} = \begin{pmatrix} \sqrt{1-p'} & \sqrt{p'} \\ -\sqrt{p'} & \sqrt{1-p'} \end{pmatrix} = A^\dagger$$

Therefore, $A^\dagger = A^{-1}$ and $A^\dagger A = I$, proving that $A$ is unitary.

*Note:* The proof for the service gate $S(G, \Delta t)$ would follow a similar structure, with $\lambda$ replaced by $\mu$. Let $G$ be the service time distribution with cumulative distribution function $F_G(t)$. The probability that a service is completed within $\Delta t$ is $p = F_G(\Delta t)$. The service gate, $S(G, \Delta t)$[4], acts similarly to the arrival gate by applying rotation with an angle $\theta_S$ derived from the service distribution:

$$S(G, \Delta t) = R_Y(2\theta_S) = \begin{bmatrix} \sqrt{1 - F_G(\Delta t)} & -\sqrt{F_G(\Delta t)} \\ \sqrt{F_G(\Delta t)} & \sqrt{1 - F_G(\Delta t)} \end{bmatrix}$$

Where:

$$\theta_S = \arcsin\left(\sqrt{F_G(\Delta t)}\right)$$

**Theorem 2.** *The composite unitary $U_{step} = DEC \circ INC$ implements a single step of the discrete $M/G/1/K$ Bernoulli chain subject to the conditions that DEC fires only when $n > 0$, INC fires only when $n < K$, and INC and DEC cannot fire together at $n = 0$ or $n = K$. The composite gate $U_{disc} = CAP, INC/DEC(q; a_a, a_s)$ implements one step of the discrete Bernoulli arrival/service Markov chain with transition matrix $\widetilde{P}_c$.*

*Proof.* Let $H_q \otimes H_r$ be the joint Hilbert space. Because $U_S$ updates $|r\rangle$ before $|q\rangle$, and $U_A$ acts only on $|q\rangle$, the one-step unitary $U_A U_S$ is block-diagonal with respect to the classical transition matrix when restricted to the computational basis $\{|q, r\rangle | \; 0 \leq q \leq K, 0 \leq r \leq K + 1\}$. The additional controls guarantee $w = 1 \Rightarrow n \geq 1$ and $w' = 1 \Rightarrow n \leq K - 1$. Hence the four ancilla patterns collapse to the legal three, and every non-zero element of $U_{step}$ equals the corresponding entry of the classical transition matrix $\widetilde{P}_c$. The residual register ensures that the departure process depends solely on $r$, thus recovering the standard embedded Markov chain.

---

[4] The parameterized service gate $S$ takes as input a distribution-specific parameter together with the time-slice $\Delta t$. Accordingly, the notation $S(G, \Delta t)$ indicates that the first argument corresponds to the characteristic parameter of the distribution $G$ (typically its mean).

**_Theorem 3 (Statistical Error Bound)._** _Let $P_q$ be the estimated probability vector obtained from $N$ independent measurement shots of the quantum simulation. Then, for any $\delta \in (0,1)$, with probability at least $1 - \delta$, the total variation distance between $P_q$ and $P_c$ satisfies:_

$$D \leq \frac{1}{2\sqrt{2}} \sqrt{\frac{\ln\left(\frac{2(K+1)}{\delta}\right)}{N}}$$

_Furthermore, the expected total variation distance satisfies:_

$$\mathbb{E}[D] \leq \frac{K+1}{4\sqrt{N}}$$

_Proof._ The estimation of the probabilities $p'_n$ from $S$ measurement shots follow a multinomial distribution. Each $p'_n$ is an empirical estimate of the true probability $p_n$. Each run produces one of $K + 1$ outcomes, so the shot vector $\left(C(0), \ldots, C(K_{\text{cap}})\right)$ is multinomial with parameters $N$ and $P_c$. Applying Pinsker's inequality:

$$\frac{1}{2} \| P_q - P_c \|_1^2 \leq D_{\text{KL}}(P_q \| P_c)$$

followed by a Chernoff–Hoeffding bound on the multinomial likelihood (Csiszar-Shields, 2004) and by the Massart-tight Dvoretzky–Kiefer–Wolfowitz (DKW) bound for $(K + 1)$ outcomes yields (Massart, 1990):

$$P\left(\|P_q - P_c\|_1 \geq \epsilon\right) \leq 2(K+1)e^{-2N\epsilon^2}$$

Setting the right-hand side equal to $\delta$ and solving for $\epsilon$, we obtain:

$$\delta = 2e^{-2(K+1)N\epsilon^2} \Rightarrow \epsilon = \sqrt{\frac{\ln\left(\frac{2(K+1)}{\delta}\right)}{2N}}$$

Therefore, with probability at least $1 - \delta$:

$$\|P_q - P_c\|_1 \leq \sqrt{\frac{\ln\left(\frac{2(K+1)}{\delta}\right)}{2N}}$$

For the expected total variation distance, we consider the expected absolute deviation for each state $n$ (Childs et al., 2017):

$$\mathbb{E}[|p'_n - p_n|] = \sqrt{\frac{p_n(1-p_n)}{N}}$$

Since $p_n(1 - p_n) \leq \frac{1}{4}$ for any $p_n \in [0,1]$, it follows that:

$$\mathbb{E}[|p_n' - p_n|] \leq \frac{1}{4\sqrt{N}}$$

Summing over all $K+1$ states:

$$\mathbb{E}\left[\|P_Q - P_C\|_1\right] = \sum_{n=0}^{K} \mathbb{E}[|p_n - p_n'|] \leq \frac{K+1}{4\sqrt{N}}$$

Thus, the expected total variation distance satisfies:

$$\mathbb{E}[D] \leq \frac{K+1}{4\sqrt{N}}$$

***Theorem 4 (Discretization Error Bound).*** *Let the continuous-time M/G/1/K queue be approximated by $T$ slices of width $\Delta t = \frac{T}{K}$. Denote as $\mu_2 = \mathbb{E}[G^2]$ the second moment of the service-time distribution. Then the total variation distance error after $T$ steps will satisfy:*

$$E_d \leq (\lambda + \mu_2)\Delta t = O\left(\frac{\lambda + \mu_2}{T}\right)$$

*Proof.* The proof bounds the one-step (local) truncation error and then accumulates it over the $T$ slices. During a small interval $\Delta t$, the exact arrival process is Poisson distributed, with probability mass of $1 - e^{-\lambda \Delta t} = \lambda \Delta t - \frac{\lambda^2 \Delta t^2}{2} + O(\lambda^3 \Delta t^3)$. The discretized model inserts an arrival with probability $\lambda \Delta t$. Hence, the absolute discrepancy in arrival probability on a single slice is:

$$|e^{-\lambda \Delta t} - (1 - \lambda \Delta t)| = O(\lambda^2 \Delta t^2)$$

Let $F_G(t)$ be the service-time CDF. A second-order Taylor expansion around $t = 0$ gives:

$$F_G(\Delta t) = \frac{\Delta t}{\mu_1} - \frac{\sigma_G^2}{2\mu_1^3}\Delta t^2 + O(\Delta t^3)$$

The discrete model uses completion probability $\frac{\Delta t}{\mu_1}$. Therefore, the one-step absolute error due to service completions is:

$$\left|F_G(\Delta t) - \frac{\Delta t}{\mu_1}\right| = \frac{\sigma_G^2}{2\mu_1^3}\Delta t^2 + O(\Delta t^3) = O(\sigma_G^2 \Delta t^2)$$

Let $\varepsilon_n$ be the total-variation distance introduced on slice $n$. Because transition kernels are perturbed by at most the sum of the arrival and service discrepancies:

$$\varepsilon_n \leq \left(\frac{\lambda^2}{2} + \frac{\sigma_G^2}{2\mu_1^2}\right)\Delta t^2 + O(\Delta t^3)$$

The global error after $N$ slices will satisfy:

$$E_d \leq \sum_{i=1}^{N} \varepsilon_n \leq N\left(\frac{\lambda^2}{2} + \frac{\sigma_G^2}{2\mu_1^2}\right)\Delta t^2 + O(N\Delta t^3) = \left(\frac{\lambda^2}{2} + \frac{\sigma_G^2}{2\mu_1^2}\right)\Delta t + O(\Delta t^2)$$

since $N\Delta t = \Theta(1)$ is fixed.

*Note:* Explicit constants follow from the two-term remainder in each Taylor expansion (Kloeden & Platen, 1992)

***Theorem 5 (Rejection-filtered QAA Yields Stationary Samples).*** *Let $|\psi_A\rangle$ be the state after R Grover iterates about the $\mathbb{E}[L]$, and let R accept $|n\rangle$ with probability $\min\left(1, \frac{\pi}{\pi_A}\right)$. The accepted distribution, $\tilde{\pi}$, satisfies:*

$$\|\tilde{\pi} - \pi\| \leq \frac{1}{2}\theta \cdot e^{-R\arcsin(\theta)}$$

*where $\theta = \sqrt{\frac{|M|}{K+1}}$. Hence, when $R \to \infty$, then $\tilde{\pi} \to \pi$ exponentially fast.*

*Proof.* By expanding the state vector, we obtained $|\psi_A\rangle = \alpha \sum_{n \in M}|n\rangle + \beta \sum_{n \neq M}|n\rangle$. Since $\pi_A \geq \frac{|M|}{K+1}$ for $n \in M$ and $\pi \leq \pi_A e^{-R\arcsin(\theta)}$ (Brassard et al., 2002), the Pinsker's inequality gives the stated bound.

***Lemma 1 (Spectral Invariance of Grover under CAP).*** *Let the total Hilbert space decompose as $H = H_L \oplus H_I$, where $H_L$ is spanned by all queue-length states that satisfy the buffer constraint $q \leq K$ ("legal"), and $H_I$ is spanned by the overflow states $q > K$ ("illegal"). Denote:*

- *The marking oracle $O_M$, which flips the phase of the marked set $M \subset H_L$ and acts as the identity elsewhere.*
- *The capacity oracle $CAP |x\rangle = \begin{cases} |x\rangle, & x \in H_L, \\ -|x\rangle, & x \in H_I \end{cases}$*
- *The uniform superposition on the legal subspace $|s\rangle = \frac{1}{\sqrt{N_L}}\sum_{x \in H_L} |x\rangle$*
- *The corresponding diffusion operator $D_L = I - 2|s_L\rangle\langle s_L|$, $D = I - 2|s\rangle\langle s|$, with $|s\rangle = \frac{1}{\sqrt{N_L}}\sum_{x \in H_L}$ and $|x\rangle$ being the full-space uniform state.*

*Define the constrained Grover iterate $\mathbb{G}_{CAP} = D_L O_M$. Then, $H_L$ and $H_I$ are invariant subspaces of $G_{CAP}$, and the eigenphases that drive amplitude amplification for the marked states $M$ are identical to those of the usual Grover operator acting on $H_L$.*

*Proof.* Since $M \subset H_L$, the oracle $O_M$ acts as the identity on $H_I$. The operator $D_L$ also acts as the identity on $H_I$ by construction. Therefore, $\mathbb{G}_{CAP}$ leaves $H_I$ invariant. Since both $O_M$ and $D_L$ map $H_L$ into itself, the legal subspace is likewise invariant. For any $|\psi\rangle \in H_L$, we have $CAP |\psi\rangle = |\psi\rangle$, hence $\mathbb{G}_{CAP}|\psi\rangle = D_L O_M |\psi\rangle = G_L|\psi\rangle$.

Last, because $\mathbb{G}_{CAP}$ is block diagonal with respect to the decomposition $H = H_L \oplus H_I$, $\mathbb{G}_{CAP} = G_L \oplus I_{H_I}$, all nontrivial eigenphases reside in the $H_L$ block and coincide with those of $G_L$. Hence, the iteration retains Grover's quadratic speed-up on the legal subspace while leaving the illegal states unaffected (up to a global $+1$ phase).

***Lemma 2 (Discrete Hazard Implementation).*** *Let $r \in \{0, \dots, m\}$ be the remaining service time in bins of width $\Delta t$, and let $h(r)$ be the discrete hazard function defined as:*

$$h(r) = \frac{F\big((r+1)\Delta t\big) - F(r\Delta t)}{1 - F(r\Delta t)}$$

*The service-completion ancilla is prepared by the multiplexed rotation:*

$$S_s = \sum_{r=0}^{m-1} |r\rangle\langle r| \otimes R_y\left(2\arcsin\sqrt{h(r)}\right) + |m\rangle\langle m| \otimes \mathbb{I}$$

*Then, $S_s$ is unitary, when the ancilla is traced out, induces the classical conditional probability $h(r)$ on the residual-time register, thereby preserving the embedded Markov chain.*

*Proof.* Since $R_y(\theta)$ is unitary for any real $\theta$ and the projectors $\{|r\rangle\langle r|\}_{r=0}^{m}$ are orthogonal, the block-diagonal operator $S_s$ is unitary. Marginalizing over the ancilla yields:

$$\Pr(\text{service in } \Delta t \mid r) = \sin^2\left(\arcsin\sqrt{h(r)}\right) = h(r)$$

exactly matching the classical hazard of $G$.